\documentclass[aps,prl,reprint]{revtex4-1}
\usepackage{amsmath}
\usepackage{amsfonts}
\usepackage{amssymb}
\usepackage{graphicx}
\usepackage{array}
\begin{document}
\title{Thermopower-based hot electron thermometry of helium surface states at 1.6 K}
%
\author {Ethan I. Kleinbaum}
\affiliation{Department of Electrical Engineering, Princeton University, Princeton, NJ 08544}
\author{Stephen A. Lyon}
\affiliation{Department of Electrical Engineering, Princeton University, Princeton, NJ 08544}
\date{\today}
\begin{abstract}
We have developed a method to probe the temperature of surface state electrons (SSE) above a superfluid Helium-4 surface using the Seebeck effect. In contrast to previously used SSE thermometry, this technique does not require detailed knowledge of the non-linear mobility.  We demonstrate the use of this method by measuring energy relaxation of SSE at 1.6 K in a microchannel device with $0.6\mu m$ deep helium. In this regime, both vapor atom scattering and 2-ripplon scattering contribute to energy relaxation to which we compare our measurements. We conclude that this technique provides a reliable measure of electron temperature while requiring a less detailed understanding of the electron interactions with the environment than previously utilized thermometry techniques.
\end{abstract}
\maketitle
Surface state electrons (SSE) above superfluid Helium-4 constitute a remarkable non-degenerate two-dimenstional electron gas (2DEG) \cite{andrei97,kono04}. These SSE float in the vacuum $\sim11$nm above the superfluid surface due to a confining potential formed by an attractive image charge in the helium and a repulsive barrier at the surface. SSE exhibit exceptional isolation from the environment. Elastic scattering processes are often used to describe the interaction of SSE with the environment using the electron mobility, which can reach values exceeding $10^8 \;\mbox{cm}^2/\mbox{Vs}$ \cite{shirahama95}. Of equal importance are the inelastic scattering processes which are characterized with the energy relaxation of SSE.

There is a great deal of interest in understanding inelastic scattering of the SSE. Inelastic processes \cite{saitoh78,dykman03,monarkha78,vilk89} are fundamental for understanding and describing a wide variety of SSE phenomena including non-linear transport\cite{saitoh78,bridges77,syvokon07,nasyedkin09,monarkha07} and microwave absorption line shapes\cite{penning00,edelman77,ando78,konstantinov08}. Further, renewed interest in these processes has emerged with the realization that the energy relaxation rates determine the coherence times of Rydberg state based SSE qubits \cite{platzman99,dykman03, monarkha06, sokolov08, monarkha07-2, monarkha10}. 

Electron thermometry is crucial for experimental measurement of energy relaxation, but thermometry of hot SSE presents a serious challenge. The lack of ohmic contacts and the exceptionally low densities preclude the use of many electron thermometry techniques developed for solid state systems \cite{giazotto06}. Instead, a common measure of the electron temperature, $T_e$, has relied on the non-linear mobility of SSE \cite{syvokon07,nasyedkin09,collin02,konstantinov07, konstantinov09,badrutdinov13,konstantinov12}.  While this approach has proven fruitful, the relationship between the mobility and $T_e$ can be complex and is known only under limited experimental conditions.  While other approaches to electron thermometry have been demonstrated for SSE, they are either confined to the Wigner crystal regime \cite{glattli88} or unable to measure the temperature of hot electrons \cite{rousseau07}.

In this paper, we describe and demonstrate the use of the Seebeck effect to measure the temperature of hot SSE in a helium microchannel device.  With the known thermopower of a non-degenerate 2DEG, we show that density measurements of a locally heated region of SSE can be related to a change in electron temperature.  Following a description and characterization of the microchannel device, we present measurements of electron heating at a helium bath temperature of 1.6K.  We find that these temperature measurements closely follow the predictions for energy relaxation which include contributions from both vapor atom and 2-ripplon scattering.  We conclude that the Seebeck effect is an effective and flexible means by which to measure the temperature of hot SSE. 


\begin{figure}[t]
\includegraphics[width=.5\textwidth]{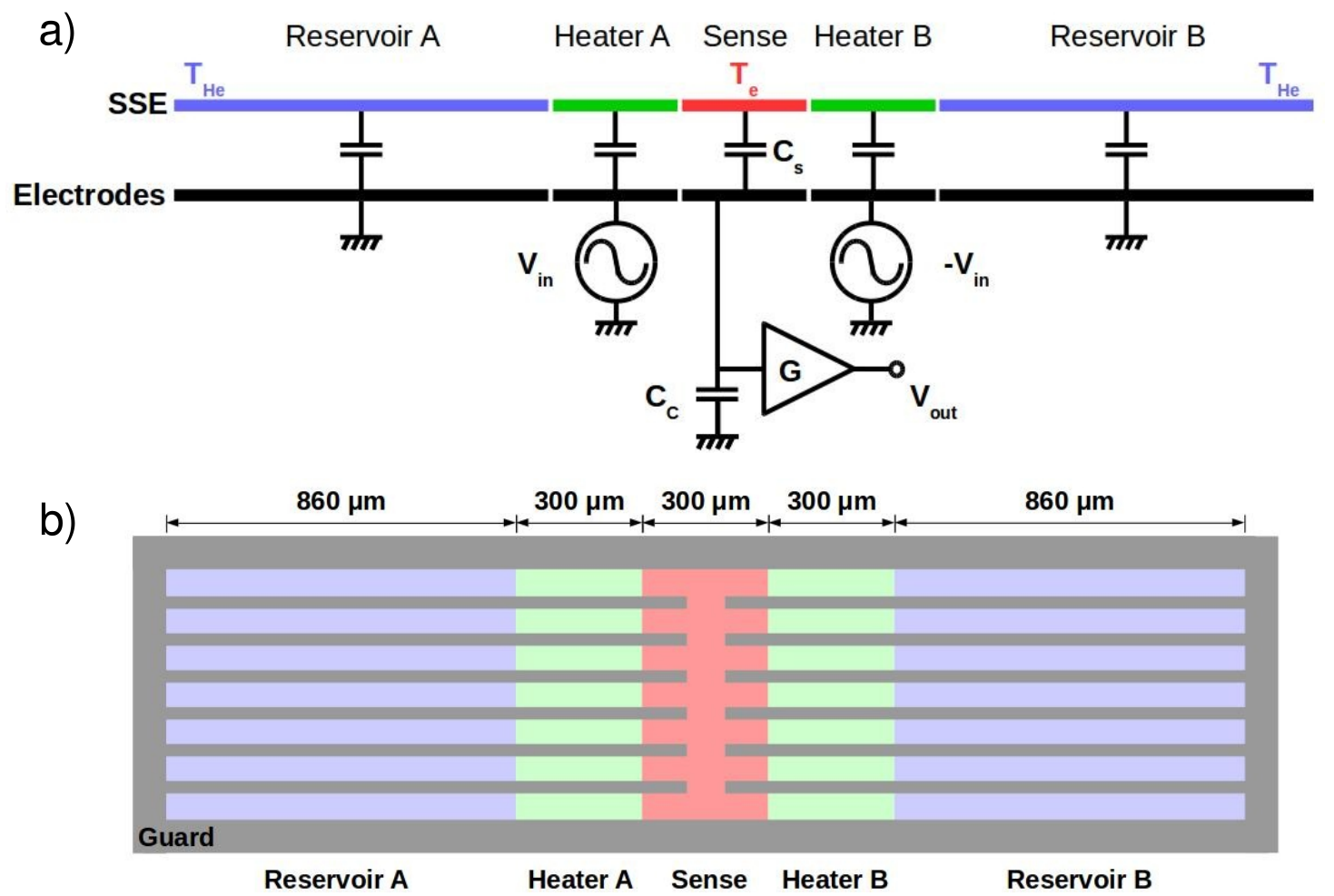}
\caption{a) A diagram of the circuit used to perform hot electron thermometry using the Seebeck effect as described in the text. (b) A diagram of the sample electrodes. Adjacent electrodes are seperated by a gap of $\sim0.5\:\mu\mbox{m}$.  The guard is isolated from the underlying electrodes with a $0.5\:\mu\mbox{m}$ layer of hardbaked photoresist.}
\end{figure}
It is well known that the Seebeck effect describes a voltage, $V_S$, which develops due to a temperature gradient, $\Delta T$, across a conductor with $V_S= Q \Delta T$ where $Q$ is referred to as the thermopower. In traditional solid state systems, the thermal voltage across a 2DEG can be measured with two voltage probes at different temperatures.  For SSE, direct measurements of electronic potentials are complicated by the necessary capacitive contact to the 2DEG. Instead, we use a new method, diagrammed in Fig.1a. The SSE layer consists of a region of locally heated electrons of temperature $T_e$ in contact with a large reservoir containing electrons thermalized to the helium bath temperature $T_{He}$. The Seebeck voltage develops between these two regions.  If the area of the reservoir is much larger than the hot electron region, then there is a well defined change in the number of electrons in the hot region of $\Delta N_e=\frac{C_{s}}{e}Q(T_e-T_{He})$ where $C_{s}$ is the capacitance between the SSE in the hot electron region and the underlying electrode in that region, referred to as the sense electrode. With the sense electrode connected to a voltage preamplifier of gain $G$ with a cable capacitance of $C_c$, the output voltage of the preamplifier will be
\begin{equation}
V_{out}=G\frac{\Delta N_e}{C_c}=G\frac{C_{s}}{C_c}Q(T_e-T_{He})
\end{equation}
$T_{He}$, $G$, $C_s$, and $C_c$ are all measurable and so, with a known value of $Q$, Eq.1 provides a direct means by which to measure the hot electron temperature.

\begin{figure}[b]
\includegraphics[width=.5\textwidth]{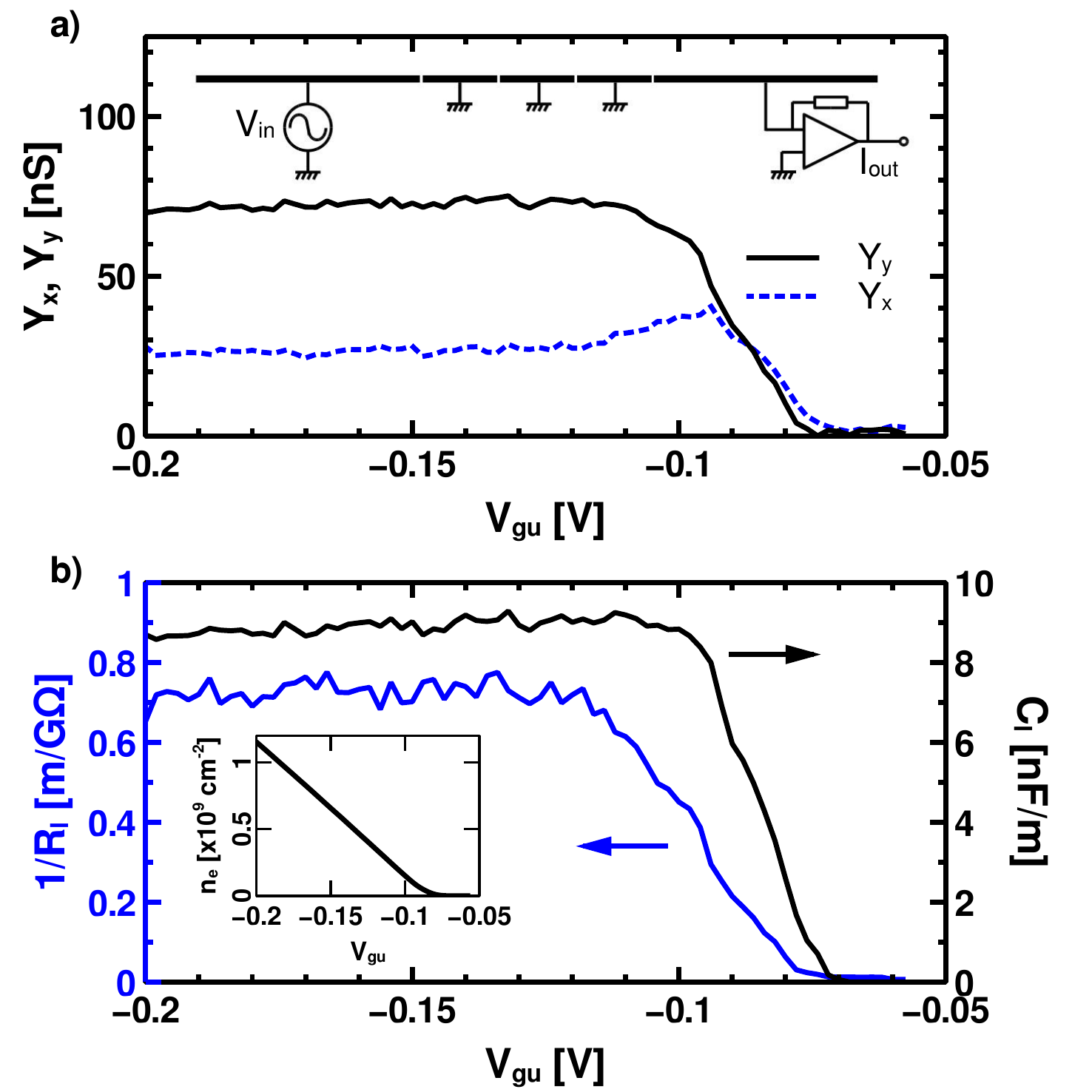}
\caption{(a) The measured complex admittance, $Y=Y_x+iY_y$, of the device at 1.6 K. (inset) A diagram of the measurement circuit. The thick black lines represent the electrodes on the bottom of the device. (b) The values of $C_l$ and $1/R_l$ calculated from the data in (a). (inset) The calculated density as a function of $V_{gu}$}
\end{figure}
Measuring $Q$ for SSE above a helium film presents a number of challenges.  Instead, we rely on the fact that for a non-degenerate 2DEG, the thermopower can be expressed generally as $Q=-\frac{k_B}{e}(\beta/k_B T_e-2-r)$ \cite{hung15} where $k_B$ is the Boltzmann constant $e$ is the charge of an electron, $\beta$ is the chemical potential and $r$ is the characteristic exponent of the energy, $E$, dependent momentum relaxation rate, $\tau(E)=\tau_0 E^r$. In the temperature and density regimes of the experiments in this paper, it is well established that $\beta/k_B T_e\approx0$ and $r\approx0$ and so the thermopower simplifies to $Q\approx 2\frac{k_B}{e}\approx172\mu V/K$. This considerably large value of $Q$ ensures that a measurable signal should be generated by minor thermal gradients.


To perform this measurement, we use a microchannel device \cite{marty86} diagrammed in Fig.1b.  Transport is confined to 70 parallel $8\;\mu\mbox{m}$ wide channels all connected by an additional microchannel traversing the center of the device. The device sits $\sim0.5\:\mbox{mm}$ above the surface of bulk helium and the channels fill by the capillary action of Helium-4. The channel geometry is defined in a $0.5\;\mu\mbox{m}$ thick layer of hard baked photoresist which is covered by a $0.1\;\mu\mbox{m}$ thick guard electrode of sputtered, metallic Niobium-Silicon (NbSi) \cite{bishop85} yielding $0.6\;\mu\mbox{m}$ deep channels. Beneath the microchannels, five additional NbSi electrodes denoted reservoir A and B, heater A and B and sense are defined above a sapphire substrate, shown in blue, green, and red, respectively, in Fig.1b. The length of the reservoir electrodes, $l_r=860\;\mu\mbox{m}$ and the lengths of the heater and sense electrodes, $l_h$ and $l_s$ respectively, are $300\;\mu\mbox{m}$. Adjacent electrodes are separated by a gap of $0.5\;\mu\mbox{m}$.
%

Much of the electrical transport in this device can be characterized from the resistance per unit length, $R_l$, of the SSE, and the capacitance per unit length, $C_l$ of the SSE to the underlying electrodes.  We begin by calculating these quantities from measurements of the device admittance, $Y=Y_x+iY_y$, using the circuit shown in the inset of Fig.2a with the transmission line equations of a rectangular microstrip \cite{mehrotra87}.  A $5\;\mbox{kHz}$ AC voltage of amplitude $3\;\mbox{mV}$ is applied to reservoir A and the induced current is detected with a current amplifier connected to reservoir B with the heater and sense electrodes grounded at $0\;\mbox{V}$. At a helium bath temperature of $T_{He}=1.6\;\mbox{K}$, electrons are emitted from a filament with the dc bias of the guard $V_{gu}$ set to $-0.2\;\mbox{V}$ until the film is saturated as measured from the dc voltage applied to the sense electrode needed to deplete the above SSE \cite{glasson00}. We measure the density dependent admittance by sweeping the dc bias of the guard to $V_{gu}=-0.06V$ thereby removing electrons from the microchannels and moving them to the thin helium film above the guard electrode \cite{rees12} where they are effectively localized. The results are shown in Fig.2a.


In Fig.2b we show the values of $C_l$ and $1/R_l$calculated numerically from the real and imaginary components of the admittance. For guard voltages below $-0.1\;\mbox{V}$, $C_l\approx9 \;\mbox{nF/m}$ in close agreement with expectations from the geometry, $C_l=\epsilon\epsilon_0 \frac{ 70 \times 8\mu m}{0.6\mu m}\approx 8.7\;\mbox{nF/m}$ where $\epsilon_0$ is the permittivity of free space and $\epsilon$ is the relative dielectric constant of helium.  Above $V_{gu}=-0.1\;\mbox{V}$, the capacitance vanishes abruptly as the charged area of the microchannels is reduced while the SSEs are depleted.

The values of $C_l$ are of utility to calculate several important parameters.  In particular, we can use $C_l$ to calculate the capacitances of the SSE to underlying electrodes.  Specifically, $C_s$ of Eq.1 is $l_s C_l$.  Further, with the assumption that the SSE potential is defined by $V_{gu}$ \cite{rees12}, the density, $n_e$ of SSE on the device can be calculated from the integral of these capacitance measurements and the geometry of the channels.  The values of $n_e$ are shown in the inset of Fig.2b.

\begin{figure}[b]
\includegraphics[width=.5\textwidth]{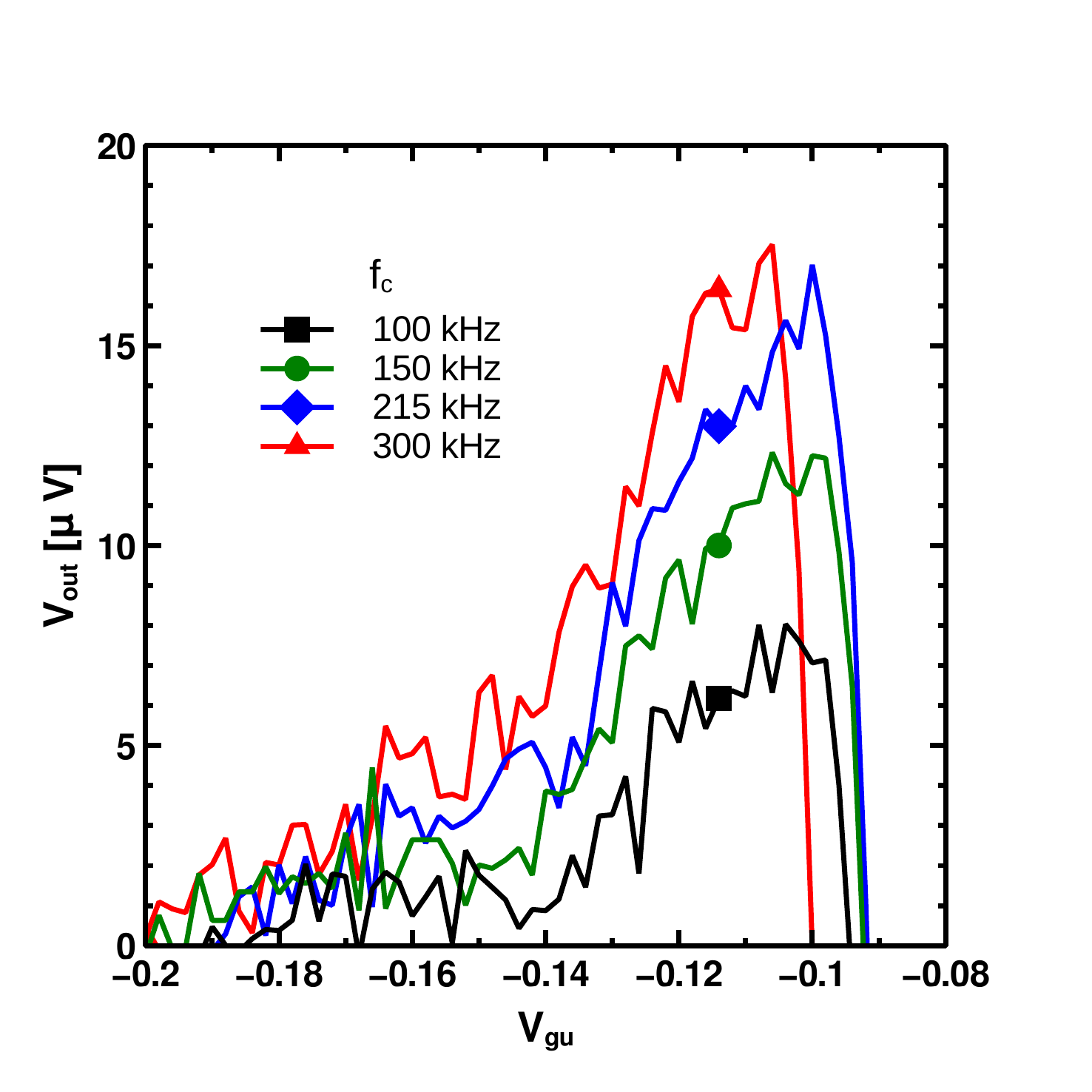}
\caption{$V_{out}$ measured at $f_{mod}=3.6\mbox{kHz}$ vs $V_{gu}$ while applying the ac modulated signal to the heaters. The measurement is repeated for $f_c$ of 100 kHz, 150 kHz, 215 kHz, and 300 kHz.}
\end{figure}
In Fig.2b, we also plot $1/R_l$ of the electron layer. $1/R_l$ shows two distinct behaviors separated at $V_{gu}=-0.12V$.  For $V_{gu}>-0.12\;\mbox{V}$, $1/R_l$ grows linearly as $V_{gu}$ becomes more negative. In the region $V_{gu}<-0.12\;\mbox{V}$, $1/R_l$ is a constant value of $\sim0.75\:\mbox{m/G}\Omega$.  The linear behavior of $1/R_l$ for $V_{gu}>-0.12\;\mbox{V}$ reflects the increase of SSE density in the microchannels as $V_{gu}$ becomes more negative leading to an increasing conductance.  At more negative $V_{gu}$ one might anticipate that $1/R_l$ should continue to increase and as such, the near constant value of the conductance below $V_{gu}=-0.12\;\mbox{V}$ is unexpected. We suggest that the constant value of $1/R_l$ reflects a density dependent suppression of the mobility. Using the values of $n_e$ determined above, at $V_{gu}=-0.12\;\mbox{V}$ the areal density is $n_e=3.5\times10^8\mbox{cm}^{-2}$ and the mobility, $\mu=22\times10^3\;\mbox{cm}^2/\mbox{Vs}$ and at $V_{gu}=-0.2\;\mbox{V}$, $n_e=1.1\times10^9\mbox{cm}^{-2}$ and $\mu=7.5\times10^3\;\mbox{cm}^2/\mbox{Vs}$. A similar density dependent $\mu$ has been measured for electrons on bulk helium \cite{mehrotra84}. 

We now turn our attention to the primary objective of this work, measuring electron heating.  We use the circuit drawn in the schematic in the inset of Fig.1a.  To heat the electrons above the helium bath temperature, we apply an amplitude modulated waveform $v_{in,a(b)}(t)=\frac{V_{0,a(b)}}{2}\sin(2\pi f_ct)(1+\sin(2\pi f_mt))$ to both heater a and b with $V_{0,a}=-V_{0,b}=12.5\mbox{mV}$.  The electrons are driven by the high frequency component of the heating voltage ($f_c$) and their average temperature is modulated at the frequency, $f_m=3.6\mbox{kHz}$. For $f_c\gg f_m$, this setup generates the time dependent resistive heating in the SSE above the sense electrode of $P_{in}(t)=I(t)^2 R_{s}=(\frac{I_0}{2}\sin(2\pi f_ct)(1+\sin(2\pi f_mt)))^2 R_{s}$ where $R_{s}=R_l l_s$ is the resistance of the 2DEG across the sense gate and
\begin{multline}
I_0=\frac{(V_{0,a}-V_{0,b})\sinh(\gamma(l_d/2))}{ Z_c \sinh(\gamma l_d)} \\ \times \left[\sinh(\gamma (l_r+l_h))-\sinh(\gamma l_r)\right]
\end{multline}
is the amplitude of the current passing through the 2DEG at the center of the sense gate with $l_d=2l_r+2l_h+l_s$, $Z_c=(i2\pi f_c C_l/R_l)^{-1/2}$ and $\gamma=(i2\pi f_c C_l R_l)^{1/2}$. Because of the symmetric geometry of the heater gates, the magnitude of the current remains nearly constant across the entire sense region allowing us to accurately estimate $P_{in}$ while only evaluating $I_0$ at a single point.
The resultant hot electron temperature signal at the sense gate is brought to a room temperature voltage amplifier of gain $G=30$ using a cable with a measured capacitance of $C_c=147\;\mbox{pF}$.  



The measurement of electron heating is initiated by emitting electrons at  $V_{gu}=-0.2\;\mbox{V}$. At the densities associated with this guard voltage, we assume electrons are well thermalized to the helium bath temperature and thus we zero offsets of $\sim10\mu \mbox{V}$ due to the non-ideal heating waveform which capacitively couples to the sense electrode. The resultant hot electron signal is measured while sweeping $V_{gu}$ until the microchannels are depleted.  We repeat this process for several values of $f_c$ and show the results in Fig.3.  For all values of $f_c$, $V_{out}$ increases at low SSE densities. Near $V_{gu}=-0.12\:\mbox{V}$, the increase in $V_{out}$ slows, then abruptly drops between $V_{gu}=-0.1\:\mbox{V}$ and $-0.09\:\mbox{V}$ with the depletion of the 2DEG. The separate measurements show qualitatively similar behaviors, though the magnitude of $V_{out}$ increases with increasing $f_c$.  The qualitative features of $V_{out}$ are consistent with the earlier characterization of the microchannel device with the inclusion of a $-22\;\mbox{mV}$ shift in $V_{gu}$ which developed after thermal cycling between measurments.  

Finally, with these values of $V_{out}$ we can use Eq.1 for precise, quantitative thermometry. In Fig.4, we plot values of $T_e(f_m)-T_{He}$ extracted from the measurements in Fig.3 against the amplitude of the power per electron applied at the modulation frequency $P(f_m)/N_e=I_0^2R_s/4$. In the same plot, we include the predictions from the models of vapor atom scattering\cite{saitoh78} (G) and 2-ripplon scattering\cite{vilk89} (2r)  and their sum.   These results are obtained by inserting $P_{in}(t)$ into the seperate energy relaxation models to determine $T_e(t)-T_{He}$ and then numerically calculating the Fourier component of the result at $f_m$ yielding $T_e(f_m)-T_{He}$. Though the exact theoretical treatment of 2-ripplon scattering remains an open question, we follow Ref.\cite{vilk89} as it has previously provided results in reasonable agreement with experimental measurements\cite{syvokon07,nasyedkin09}. We simplify the analysis by assuming electrons populate only the ground state subband for vapor atom scattering and by setting the holding field $E_\perp=0\;\mbox{V/m}$ for 2-ripplon scattering.  These two simplifications alter the prediction of $T_e(f_m)-T_{He}$ by less than the uncertainty of our measurements. Finally, we estimate that for the experimental conditions of this work the electronic thermal conductance of the 2DEG from Wiedemann-Franz Law is not significant with respect to the previously described relaxation mechanisms and thus we disregard it in our analysis.
%
%

There are several notable features of our measurements. Initially striking is the close overlap of the measurements for different values of $f_c$. The reproducibility of these measurements under different conditions indicates that the measured signal is explicitly a function of $P_{in}/N_e$ which provides confidence that our measurements are not the results of spurious sources given the non-linear relationship between $I$ and $P_{in}$.  Further, holding $P_{in}/N_e$ constant for different values of $f_c$ requires $R_l$, $C_l$ and $N_e$ to adjust in a non-trivial manner, thus demonstrating that the characterization presented above provides an accurate description of our device and that our calculation of the applied power is appropriate. 

Further confidence in the validity of these measurements is drawn from a comparison to the theoretical predictions for energy relaxation.  The qualitative features and the quantitative values of the measurements of $T_e(f_m)-T_{He}$ are in reasonable agreement with the values predicted from the sum of vapor atom scattering and 2-ripplon scattering. We emphasize that our analysis does not include any adjustable parameters which might artifically improve the comparison.

 
 
The predicted values of $T_e(f_m)-T_{He}$ are approximately $50\%$ larger than the measurements.  This discrepancy is unsurprising. The predictions used here are made for SSE on bulk helium while our measurements are performed on a microchannel device. These distinct conditions may require different treatments to satisfactorily describe the inelastic scattering in the system.  From these measurements we are unable distinguish from which source of inelastic scattering this discrepancy arises because the two contributions are of similar magnitude.  This uncertainty may be resolved in future work by performing measurements at lower temperatures at which the 2-ripplon scattering dominates.
\begin{figure}
\includegraphics[width=.5\textwidth]{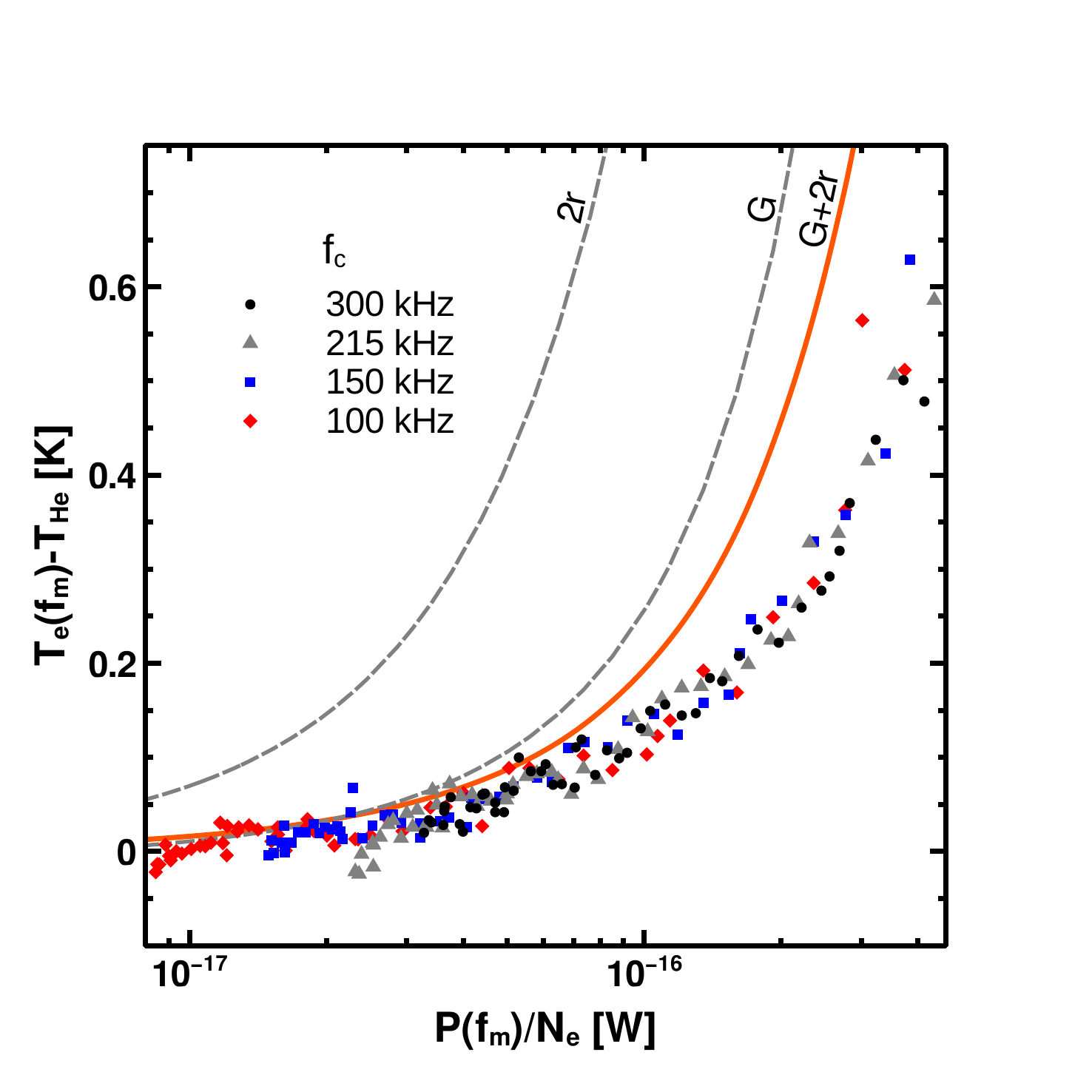}
\caption{The electron heating signal versus the applied power per electron. The symbols are values calculated from the data in Fig.3 as described in the text. The dashed lines are the separate contributions to energy relaxation of the 2-ripplon and gas atom scattering labeled as 2r and G, respectively. The solid line labeled G+2r is the sum of the two contributions. }
\end{figure}

These measurements illustrate several key advantages of decoupling electron thermometry from non-linear transport. Principally, we have probed the electron temperature without the need a detailed model of electron mobility. While the estimate of $Q$ relied on the knowledge of $r$, this is a much weaker requirement than knowing the exact functional relationship between $\mu$ and $T_e$. Of additional advantage, use of the Seebeck effect has allowed for hot electron thermometry in a regime which cannot be probed with non-linear transport. For SSE on helium at 1.6 K, the mobility does not change from the equilibrium value until electron temperatures are nearly 10 K. Nevertheless, utilizing the Seebeck effect has allowed us to probe energy relaxation at electron temperatures which are less then 1 K above the helium bath temperature.

These advantages may prove crucial in answering open questions regarding the fundamental nature of energy relaxation of SSE. In particular, there is uncertainty regarding the appropriate theoretical treatment of 2-ripplon scattering.  While  experimental results have provided initial characterization of 2-ripplon energy relaxation \cite{syvokon07,nasyedkin09,badrutdinov13}, more detailed studies under a range of experimental conditions are necessary to achieve definitive conclusions.

In summary we have developed a new technique to probe the electron temperature of SSE above superfluid helium.  Using the Seebeck effect, we have performed electron thermometry from simple voltage measurements.  We have used this technique to measure the energy relaxation of SSE on a microchannel device at 1.6 K where both vapor atom scattering and 2-ripplon scattering contribute.  We find that our measurements in approximate quantitative agreement with simplified models of electron energy loss.  These results demonstrate that the Seebeck effect provides a flexible and effective means by which to measure the hot electron temperature of SSE. 

This research was supported by the NSF (Grant No.
DMR-1506862).  Devices were fabricated in the Prince-
ton Institute for the Science and Technology of Materials
Micro/Nano  Fabrication  Laboratory.


\begin{thebibliography}{l}
\bibitem{andrei97} \textit{Two-Dimensional Electron Systems on Helium and Other Cryogenic Substrates}, edited by E.Y. Andrei (Kluwer Academic, Dordrecht, MA, 1997)
\bibitem{kono04} Y. Monarkha and K. Kono, \textit{Two-Dimensional Coulomb Liquids and Solids} (Springer-Verlag, Berlin, 2004)
\bibitem{shirahama95} K. Shirahama, S. Ito, H. Suto and K. Kono, J. Low Temp. Phys. \textbf{101}, 439 (1995)
\bibitem{saitoh78} M. Saitoh and T. Aoki, J. Phys. Soc. Jpn. \textbf{44}, 71 (1978)
\bibitem{monarkha78} Y.P. Monarkha, Fiz. Nizk. Temp \textbf{4}, 1093 (1978) [Sov. J. Low Temp. Phys. \textbf{4}, 515 (1978)]
\bibitem{dykman03} M.I. Dykman, P.M. Platzman and P. Seddighrad, Phys. Rev. B \textbf{67}, 155402 (2003)

\bibitem{vilk89} Y.M. Vil’k and Y.P. Monarkha, Fiz. Nizk. Temp. \textbf{15}, 235 (1989) [Sov. J. Low Temp. Phys. \textbf{15}, 131 (1989)]

\bibitem{bridges77} F. Bridges and J. McGill, Phys. Rev. B \textbf{15}, 1324 (1977)
\bibitem{syvokon07} V. Syvokon, Y. Monarkha, K. Nasyedkin and  S. Sokolov, J. Low Temp. Phys. \textbf{148}, 169 (2007)
\bibitem{nasyedkin09} K.A. Nasyedkin, V.E. Sivokon, Y.P. Monarkha and S.S. Sokolov, Low Temp. Phys. \textbf{35}, 757 (2009)
\bibitem{monarkha07} Y.P. Monarkha, D. Konstantinov and K. Kono, J. Phys. Soc. Jpn \textbf{76}, 0124702 (2007).






\bibitem{ando78}T. Ando, J. Phys. Soc. Jpn. \textbf{44}, 765 (1978)
\bibitem{edelman77} V.S. Edel'man, JETP Lett. \textbf{25}, 394 (1977)
\bibitem{penning00} F.C. Penning, O. Tress, H. Bluyssen, E. Teske, M. Seck, P. Wyder and V.B. Shikin, Phys. Rev. B \textbf{61}, 4530 (2000) 

\bibitem{konstantinov08} D. Konstantinov, H. Isshiki, H. Akimoto, K. Shirahama, Y. Monarkha and K. Kono, J. Phys. Soc. Jpn. \textbf{77}, 034705 (2008) 
\bibitem{monarkha06} Y.P. Monarkha and S.S. Sokolov, Low Temp. Phys. \textbf{32}, 970 (2006)
\bibitem{monarkha07-2} Y.P. Monarkha and S.S. Sokolov, J. Low Temp. Phys. \textbf{148}, 157 (2007)
\bibitem{sokolov08} S.S. Sokolov Y.P. Monarkha, J.M. Villas-Boas and Nelson Studart, Low Temp. Phys. \textbf{34}, 385 (2008)
\bibitem{monarkha10} Y.P. Monarkha, S.S. Sokolov, A.V. Smorodin and Nelson Studart, Low Temp. Phys. \textbf{36}, 565 (2010)
\bibitem{platzman99} P.M. Platzman and M.I. Dykman, Science \textbf{284}, 1967 (1999)

\bibitem{giazotto06} F. Giazotto, T.T. Heikkila, A. Luukanen, A.M. Savin and J.P. Pekola, Rev. Mod. Phys. \textbf{78}, 217 (2006)
\bibitem{collin02} E. Collin, et al., Phys. Rev. Lett. \textbf{89}, 245301 (2002)
\bibitem{konstantinov07} D. Konstantinov, H. Isshiki, Y. Monarkha, H. Akimoto, K. Shirahama and K. Kono, Phys. Rev. Lett. \textbf{98}, 235302 (2007)


\bibitem{konstantinov09} D. Konstantinov, M.I. Dykman, M.J. Lea, Y. Monarkha and K. Kono, Phys. Rev. Lett. \textbf{103}, 096801 (2009)
\bibitem{badrutdinov13} A.O. Badrutdinov, D. Konstantiov, M. Watanabe and K. Kono, EPL \textbf{104}, 47007 (2013)
\bibitem{konstantinov12} D. Konstantinov, M.I. Dykman, M.J. Lea, Y.P. Monarkha and K. Kono, Phys. Rev. B \textbf{85}, 155416 (2012)


\bibitem{glattli88} D.C. Glattli, E.Y. Andrei and F.I.B. Williams, Phys. Rev. Lett. \textbf{60}, 420 (1988)
\bibitem{rousseau07} E. Rousseau, Y. Mukharsky, D. Ponarine, O. Avenel and E. Varoquaux, J. Low Temp. Phys. \textbf{148}, 193 (2007)

\bibitem{hung15} N.T. Hung, A.R.T. Nugraha, E.H. Hasdeo, M.S. Dresselhaus and R. Saito, Phys. Rev. B \textbf{92}, 165426 (2015)



\bibitem{marty86} D. Marty, J. Phys. C \textbf{19}, 6097 (1986)

\bibitem{bishop85} D.J. Bishop, E.G. Spencer and R.C. Dynes, Sol. Stat. Elect. \textbf{28}, 73 (1985)
\bibitem{mehrotra87} R. Mehrotra and A.J. Dahm, J. Low Temp. Phys \textbf{67}, 115 (1987)
\bibitem{glasson00} P. Glasson, S.E. Andresen, G. Ensell, V. Dotsenko, W. Bailey, P. Fozooni, A. Kristensen and M.J. Lea, Physica B \textbf{284-288}, 1916 (2000)
\bibitem{rees12} D.G. Rees, I. Kuroda, C.A. Marrache-Kikuchi, M. H\"ofer, P. Leiderer and K. Kono, J. Low Temp. Phys. \textbf{166}, 107 (2012)
\bibitem{mehrotra84} R. Mehrotra, C.J. Guo, Y.Z. Ruan, D.B. Mast and A.J. Dahm, Phys. Rev. B \textbf{29}, 5239 (1984)



\end{thebibliography}
\end{document}